\documentclass[aps,prl,twocolumn,reprint]{revtex4-1}
\usepackage{amsmath}
\usepackage{amssymb}
\usepackage{graphicx}

\begin{document}
\title{(LaTiO$_3$)$_n$/(LaVO$_3$)$_n$ as a model system for unconventional charge transfer and polar metallicity}
\author{Yakui Weng}
\author{Jun-Jie Zhang}
\author{Bin Gao}
\author{Shuai Dong}
\email{Corresponding author: sdong@seu.edu.cn}
\affiliation{School of Physics, Southeast University, Nanjing 211189, China}

\date{\today}

\begin{abstract}
At interfaces between oxide materials, lattice and electronic reconstructions always play important roles in exotic phenomena. In this study, the density functional theory and maximally localized Wannier functions are employed to investigate the (LaTiO$_3$)$_n$/(LaVO$_3$)$_n$ magnetic superlattices. The electron transfer from Ti$^{3+}$ to V$^{3+}$ is predicted, which violates the intuitive band alignment based on the electronic structures of LaTiO$_3$ and LaVO$_3$. Such unconventional charge transfer quenches the magnetism of LaTiO$_3$ layer mostly and leads to metal-insulator transition in the $n=1$ superlattice when the stacking orientation is altered. In addition, the compatibility among the polar structure, ferrimagnetism, and metallicity is predicted in the $n=2$ superlattice.
\end{abstract}
\maketitle

\section{Introduction}
Transition-metal oxide heterostructures have become one of the most-concerned branches of condensed matter physics and material science, which show plenty novel physical phenomena and are probably lead the revolution of electronic devices \cite{Dagotto:Sci07,Hammerl:Sci,Takagi:Sci,Hwang:Nm}. Around the interfaces between oxides, both the lattice structure and electronic structure would be reconstructed, which co-determine emergent exotic physical properties. For example, the two-dimensional electronic gas between two insulators was found in LaAlO$_3$/SrTiO$_3$ \cite{Ohtomo:Nat04,Nakagawa:Nm}, and orientation-dependent magnetism was observed in LaNiO$_3$/LaMnO$_3$ \cite{Gibert:Nm,Dong:Prb13.1} and LaFeO$_3$/LaCrO$_3$ superlattices \cite{Ueda:Sci,Zhu:Jap}.

As an important driving force for electronic reconstruction, charge transfer, namely the electron leakage from one side to another, can modulate the local electron density and chemical potential near the interfaces. Although this effect is well-known in traditional semiconductor $p-n$ junctions, sometimes unconventional charge transfer against the simple band alignment scenario may occur in these correlated electron systems, e.g. the electron transfer from Ti$^{3+}$ to Fe$^{3+}$ in $R$FeO$_3$/$R$TiO$_3$ \cite{Kleibeuker:Prl,He:Prb16,Zhang:Prb15} and Ti$^{3+}$ to Ni$^{3+}$ in $R$NiO$_3$/GdTiO$_3$ \cite{Grisolia:Np}. In these Fe$^{3+}$/Ni$^{3+}$-Ti$^{3+}$ interfaces, the original occupied states of Ti$^{3+}$ is lower than the unoccupied states of Fe$^{3+}$/Ni$^{3+}$, but the charge transfer still happens. The possible underlying mechanisms were attributed to the ``soft" Hubbard bands, high-spin/low-spin magnetic transition, or the covalence of the metal-oxygen bond, all of which can seriously modulate the original electronic structures \cite{Kleibeuker:Prl,Zhang:Prb15,Grisolia:Np}.

As one of intriguing physical results of electronic reconstruction, the metal-insulator transition may emerge in some heterostructures like (LaMnO$_3$)$_{2n}$/(SrMnO$_3$)$_n$ superlattices \cite{Bhattacharya:Prl,Dong:Prb08.3,Aruta:Prb,Nanda:Prb09}. Understanding and controlling metal-insulator transition in oxide heterostructures can provide design rules for electronic devices.

In this work, the (LaTiO$_3$)$_n$/(LaVO$_3$)$_n$ superlattices have been studied using the density functional theory (DFT) and maximally localized Wannier functions (MLWFs). These two perovskites own identical crystalline structure (orthorhombic No. 62 $Pbnm$) and A-site ions, which provide an ideal platform to explore the charge transfer between Ti and V, without other contributions from A-site ions and complex polar discontinuity. In the following study, several interesting physical phenomena have been revealed, including the unconventional charge transfer, orientation-dependent metal-insulator transition, and polar metallicity with noncentrosymmetric structure. In this sense, such a simple combination can be a model system to illustrate the plenty physics of oxide interfaces.

\section{Models and Methods}
The ground state of LaTiO$_3$ bulk is a G-type antiferromagnetic (G-AFM) Mott insulator with GdFeO$_3$-type distortion \cite{Cwik:Prb,Hemberger:Prl03}, while LaVO$_3$ has a C-type antiferromagnetic (C-AFM) order with Jahn-Teller distortion in low temperatures \cite{Bordet:Jssc}. Their experimental lattice constants ($a$, $b$, $c$) are very close: ($5.637$, $5.619$, $7.916$) for LaTiO$_3$ \cite{Komarek:Prb} and ($5.553$, $5.553$, $7.845$) for LaVO$_3$ in unit of {\AA} \cite{Bordet:Jssc}, which ensure the promising epitaxial growth of multilayers. The widely used substrate LaGaO$_3$ ($5.524$ {\AA}, $5.492$ {\AA}, $7.774$ {\AA}) \cite{Slater:Jssc} is chosen for the appropriate in-plane compressive strain, namely the in-plane lattice constants of LaVO$_3$ and LaTiO$_3$ were fixed to match the substrate. The choice of LaGaO$_3$ can avoid the termination issue since the films and substrate share the identical A-site cation.

The following calculations are performed using Vienna {\it ab initio} Simulation Package (VASP) \cite{Kresse:Prb93,Kresse:Prb96} based on the local density approximation (LDA) method with the projector-augmented wave (PAW) potentials. The cutoff energy of the plane wave is $550$ eV. A $7\times7\times5$, $5\times7\times7$ and $7\times7\times5$ Monkhorst-Pack $k$-point mesh centered at $\varGamma$ point are adopted for (LaTiO$_3$)$_1$/(LaVO$_3$)$_1$ superlattices stacking along the [001], [110] and [111] directions respectively, while it is $6\times6\times2$ for the [001]-orientated (LaTiO$_3$)$_2$/(LaVO$_3$)$_2$ superlattice. Both the out-of-plane lattice constants and atomic positions are fully relaxed till the Hellman-Feynman forces are converged to less than $0.01$ eV/{\AA}.

The Hubbard repulsion $U_{\rm eff}$ ($=U-J$) is imposed on Ti's/V's $3d$ and La's $4f$ orbitals using the Dudarev implementation \cite{Dudarev:Prb}. According to previous literature \cite{Weng:Jap14,Fang:Prb03,Dong:Jap14}, $U_{\rm eff}$(Ti)=$2.3$ eV, $U_{\rm eff}$(V)=$3$ eV, and $U_{\rm eff}$(La)=$8$ eV are proper to reproduce the experimental properties and thus are adopted as default parameters in the following calculations, if not noted explicitly.

The MLWFs are employed to fit the DFT bands \cite{Marzari:Prb,Souza:Prb,Mostofi:Cpc}.

\section{Results and discussion}
\begin{figure}
\centering
\includegraphics[width=0.45\textwidth]{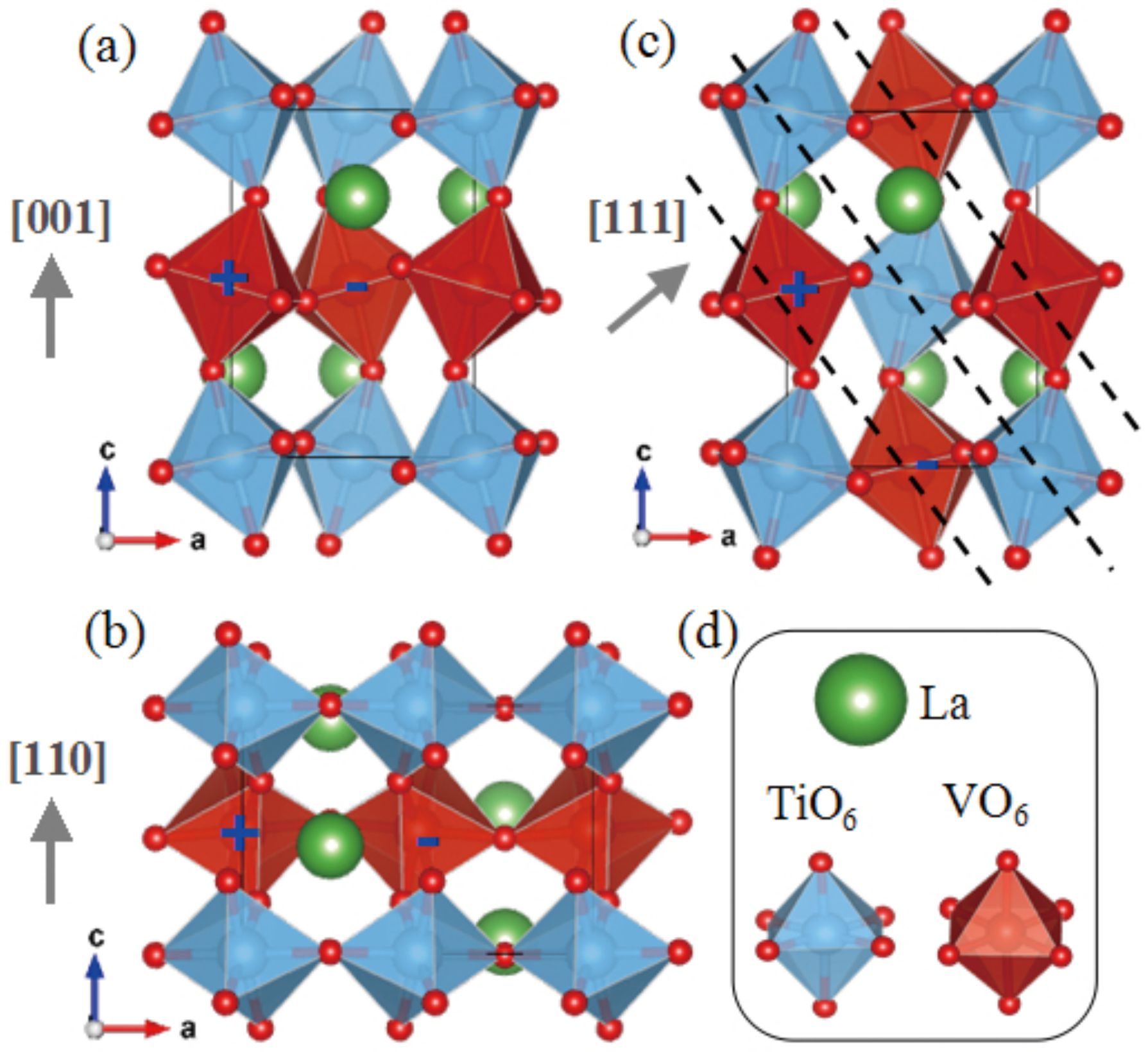}
\includegraphics[width=0.45\textwidth]{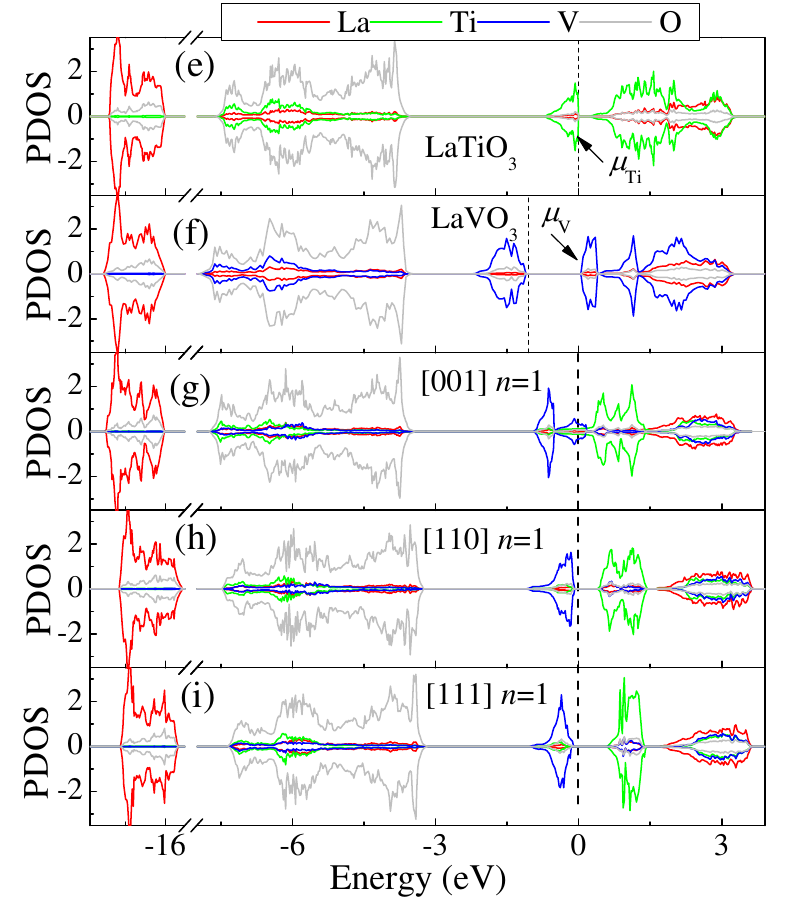}
\caption{(a) Sketch of (LaTiO$_3$)$_1$/(LaVO$_3$)$_1$ superlattice stacking along the [001] direction. $+$ and $-$ denote the spin directions. (b) The [110]-oriented superlattice. (c) A unit cell of [111]-oriented superlattice. The (111) plane is marked by (black) broken line. (d) Elements of LaTiO$_3$/LaVO$_3$. (e-f) PDOS for strained LaTiO$_3$ and LaVO$_3$. The band alignment refers to La's $5p$ bands. The Fermi level for each case is marked by a (black) broken line. (g-i) PDOS for $n=1$ superlattices along the [001], [110], and [111] direction, respectively.}
\label{F1}
\end{figure}

First, the parent materials have been checked. By comparing the energies of various magnetic orders, i.e., ferromagnetic (FM), A-type antiferromagnetic (A-AFM), C-AFM, G-AFM, our calculation confirms that the magnetic ground states of bulk LaTiO$_3$ and LaVO$_3$ are G-AFM and C-AFM, respectively. The corresponding magnetic moments of Ti and V are $\sim0.57$ $\mu_{\rm B}$ and $\sim1.75$ $\mu_{\rm B}$ respectively. Also the relaxed lattice constants are very close to the experimental values \cite{Komarek:Prb,Bordet:Jssc}. These agreements imply reliable DFT description of these two materials.

Then the strain effects from LaGaO$_3$ substrate have been studied. Upon the compressive strain, both LaTiO$_3$ and LaVO$_3$ undergo a phase transition from the original magnetic orders to A-AFM, in agreement with previous studies \cite{Weng:Jap14,Weng:Prb10}. The atomic projected density of states (PDOS) of strained LaTiO$_3$ and LaVO$_3$ are displayed in Fig.~\ref{F1}(e-f). Both materials remain insulating as in the unstrained condition, with energy gaps $\sim0.20$ eV for LaTiO$_3$ and $\sim1.15$ eV for LaVO$_3$.

\subsection{A. $1+1$ superlattice}
\subsubsection{A.1 Band alignment \& charge transfer}

For the $n=1$ superlattice, three orientations, i.e., [001], [110], and [111] (see Fig.~\ref{F1}(a-c)), have been studied. First, the possible magnetic ground states are investigated within a minimal unit cell ($2$ V plus $2$ Ti). In all cases, the local magnetic moments of Ti are quenched to near zero. Meanwhile, the corresponding V's moments are promoted to: $\sim2.1$ $\mu_{\rm B}$/V, $\sim2.3$ $\mu_{\rm B}$/V, and $\sim2.4$ $\mu_{\rm B}$/V for the three orientations respectively. These results are qualitatively independent on the preset magnetic orders. Therefore, the electron transfer from Ti to V is unambiguous in all cases, further confirmed by the PDOS (Fig.~\ref{F1}(g-i)). For all cases, the two V ions in the minimal cell are coupled antiferromagnetically.

According to Fig.~\ref{F1}(e-f), the topmost valence band of LaTiO$_3$ is from one $t_{\rm 2g}$ orbital of Ti (whose position is denoted as $\mu_{\rm Ti}$), and the lowest conducting band of LaVO$_3$ is from the spin-up $t_{\rm 2g}$ orbitals of V (whose position is denoted as $\mu_{\rm V}$). By choosing the deep energy bands of La's $5p$ orbitals as the common reference point, the band alignment can be obtained directly, whose validity can be further confirmed by the overlapped energy windows of La's $5p$ orbitals in the superlattice calculations (Fig.~\ref{F1}(g-i)). Since $\mu_{\rm Ti}$ is lower than $\mu_{\rm V}$, intuitively, the charge transfer would not occur, keeping the Ti$^{3+}$-V$^{3+}$ configuration. However, unexpected charge transfer happens in all three superlattices, as shown in Fig.~\ref{F1}(g-i), in opposite to the band alignment scenario but in agreement with aforementioned change of local moments. This result is similar to the charge transfer in ($R$TiO$_3$)$_n$/($R$FeO$_3$)$_n$ and ($R$NiO$_3$)$_n$/(GdTiO$_3$)$_n$ superlattices \cite{Zhang:Prb15,Kleibeuker:Prl,He:Prb16,Grisolia:Np}.

To understand such unconventional charge transfer, we once proposed that the Hubbard bands of Mott insulators are not rigid but fragile against the change of electron density \cite{Zhang:Prb15}. Other mechanisms, like the covalence of the metal-oxygen bond \cite{Grisolia:Np}, may be also responsible for the violation of band alignment. Here another driving force from the lattice reconstruction is evidenced. As mentioned before, LaTiO$_3$ owns a moderate GdFeO$_3$-type and Jahn-Teller distortions, which prefer the staggered $d_{xz}$/$d_{yz}$ orbital ordering  \cite{Mochizuki:Njp}. In contrast, the lattice distortions in LaVO$_3$ are much weaker. Thus, the distortions of Ti-O octahedra will be suppressed in these superlattices, which suppress the orbital ordering. Then the original low-lying $d_{xz}$/$d_{yz}$ bands will be lifted up and as a consequence the electron can leak from Ti to V once $\mu_{\rm Ti}$ is over $\mu_{\rm V}$.

Taking the [001]-oriented superlattice for example, the structural distortions are analyzed to verify above argument. First, as shown in Table~\ref{table1}, both the in-plane and out-of-plane Ti-O-Ti(V) bonds are straighter in the superlattice than the original ones in LaTiO$_3$. This suppressed GdFeO$_3$-type distortion is beneficial for $3d$-$2p$ orbital hybridization, leading to wider bandwidth.

Second, the breathing mode $Q_1$ and Jahn-Teller mode $Q_2$/$Q_3$ \cite{Dagotto:Prp,dong:prb11}, are also analyzed in Table~\ref{table1}. For the [001] case, the breathing mode $Q_1$, which characterizes the size of oxygen octahedral cage, is shrunk by $7.5$ pm in superlattice, in agreement with higher valence of Ti. The Jahn-Teller modes, which breaks the cubic symmetry and splits the degeneration of triplet $t_{\rm 2g}$ orbitals, are also significantly changed, which suppress the original $d_{xz}$/$d_{yz}$ orbital ordering but prefer the $d_{xy}$ orbital. Similar changes also occur in the [110] and [111] cases.

\begin{table}
\caption{The optimized Ti-O-Ti(V) bond angles $\alpha$ (in the $x$-$y$ plane) and $\beta$ (along the $z$-axis) for strained LaTiO$_3$, [001]-, [110]- and [111]-stacking superlattices. Here the $x$-$y$-$z$ is the coordinate for pseudocubic lattice framework. The corresponding breathing mode ($Q_1$) and Jahn-Teller modes ($Q_2$ and $Q_3$) for Ti-O are also listed. All modes are in unit of pm. Since only the change of $Q_1$ is referable, the original $Q_1$ of strained LaTiO$_3$ is set as the reference.}
\begin{tabular*}{0.48\textwidth}{@{\extracolsep{\fill}}cccccc}
\hline \hline
 & $\alpha$ ($^{\circ}$) & $\beta$ ($^{\circ}$) & $Q_1$ & $Q_2$ & $Q_3$\\
\hline
strained LaTiO$_3$ & $153.4$ & $158.6$ & $-$ & $3.5$ & $2.8$\\
$n=1$ [001] & $158.9$ & $159.3$ & $-7.5$ & $0.1$ & $-3.0$\\
$n=1$ [110] & $157.0$ & $157.0$ & $-7.8$ & $0.2$ & $2.0$\\
$n=1$ [111] & $158.1$ & $156.7$ & $-8.9$ & $0.2$ & $0.0$\\
\hline \hline
\end{tabular*}
\label{table1}
\end{table}

In addition, the reduced (effective) dimension will also affect the band structures. In bulks, each Ti (V) has six nearest-neighbor Ti (V), forming a three-dimensional network. However, in the [001] case, each Ti (V) has four nearest-neighbor Ti (V), forming a pseudo-square network of Ti (V). While in the [110] case, each Ti (V) only has two nearest-neighbor Ti (V), forming a quasi-one-dimensional Ti (V) chain. Furthermore, in the [111] case, each Ti (V) has surrounded by six nearest-neighbor V (Ti), forming a quasi-zero-dimension structure of Ti (V). In principle, the reduced dimension will narrow the bandwidth. Similar dimensional effects have been emphasized for LaMnO$_3$/LaNiO$_3$ as well as SrRuO$_3$/SrTiO$_3$ superlattices \cite{Dong:Prb13.1,Gu:Qm}.

All these factors will contribute to the electronic reconstruction, which is a result of cooperative mechanisms. Thus, for each oxide superlattice, a specific careful study is needed to reveal the charge transfer, while a rough band alignment based on information of bulks is not enough.

\subsubsection{A.2 Metal-insulator transition \& orbital ordering}
As shown in Fig.~\ref{F1}(g-i), the [001]-oriented superlattice is metallic, while the other two are insulating. This metal-insulator transition enriches the orientation-dependent physics in oxide heterostructures. Many previous works have revealed plenty orientation-dependent physics of perovskite films/superlattices, e.g., magnetism \cite{Gibert:Nm,Dong:Prb13.1,Ueda:Sci,Zhu:Jap,Huang:Jap13}, topological phase \cite{Xiao:Nc11,Weng:Prb15}, as well as superconductivity \cite{Nakagawa:Nm}. The present study adds one more interesting topic.

\begin{figure}
\centering
\includegraphics[width=0.48\textwidth]{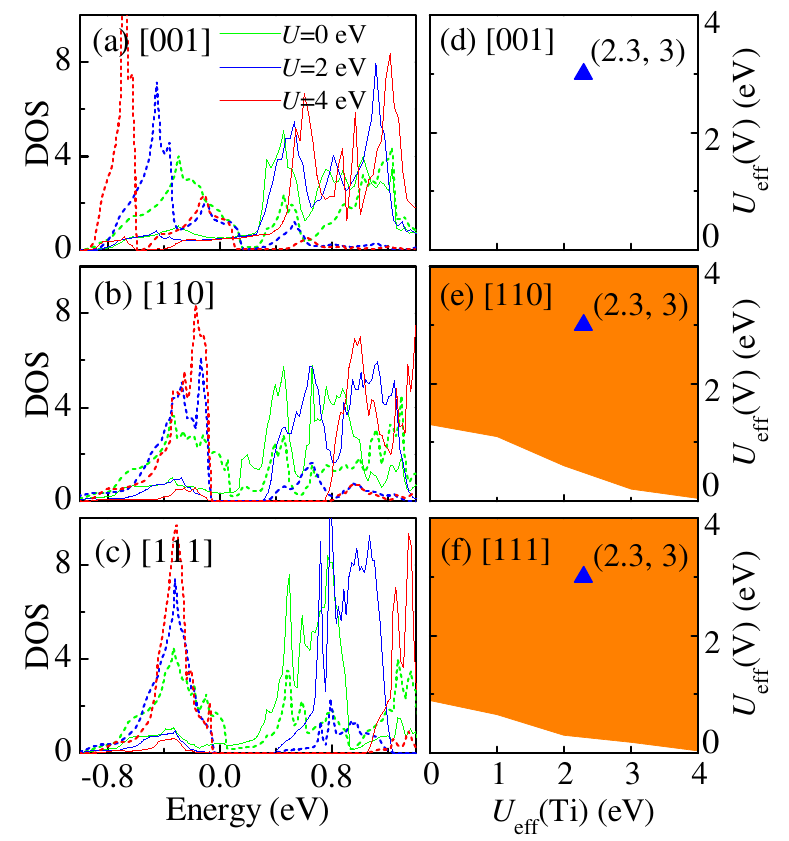}
\caption{The correlation-dependent electronic structures and phase diagrams. (a-c) PDOS of Ti (solid curves) and V (broken curves), calculated with some selected values of $U_{\rm eff}$. For simplify, here $U_{\rm eff}$(Ti)=$U_{\rm eff}$(V). (d-f) The ground state phase diagrams of (LaTiO$_3$)$_1$/(LaVO$_3$)$_1$ as a function of $U_{\rm eff}$(Ti) and $U_{\rm eff}$(V). Metallic and insulating regions are distinguished by white and orange colors.}
\label{F2}
\end{figure}

To reveal the evolution of interfacial electronic structure upon Hubbard-type correlation, here a wide parameter space of $U_{\rm eff}$ is scanned, as shown in Fig.~\ref{F2}(a-c). First, the charge transfer tendency is independent of $U_{\rm eff}$ nor orientation, i.e., electron always leaks from Ti$^{3+}$ to V$^{3+}$. This implies a robust conclusion for the unconventional charge transfer. Second, the Coulombic correlation can further enhance the band offset between Ti's $t_{\rm 2g}$ and V's $t_{\rm 2g}$, to promote the charge transfer. Third, the charge transfer in the [001]-oriented case is always incomplete, rendering a metallic interface even when $U_{\rm eff}$ is quite large. In contrast, in other two cases, the charge transfer becomes complete once moderate $U_{\rm eff}$'s are applied, leading to insulating interfaces. The metal-insulator phase diagrams are summarized as Fig.~\ref{F2}(d-f). The [001]-oriented (LaTiO$_3$)$_1$/(LaVO$_3$)$_1$ superlattice is always metallic, independent of $U_{\rm eff}$ (in a reasonable region). In contrast, for the [110] and [111] cases, the systems are metallic only when both $U_{\rm V}$ and $U_{\rm Ti}$ are very small (not reasonable for real V and Ti), but become insulating in more reasonable region. Therefore, it is safe to conclude the highly probably metal-insulator transition from the [001]-stacking superlattice to the [110]- or [111]-stacking superlattice. As stated in the method section, $U_{\rm eff}$(Ti)=$2.3$ eV and $U_{\rm eff}$(V)=$3$ eV are the best choice, which will be fixed in the following calculations.

\begin{figure}
\centering
\includegraphics[width=0.45\textwidth]{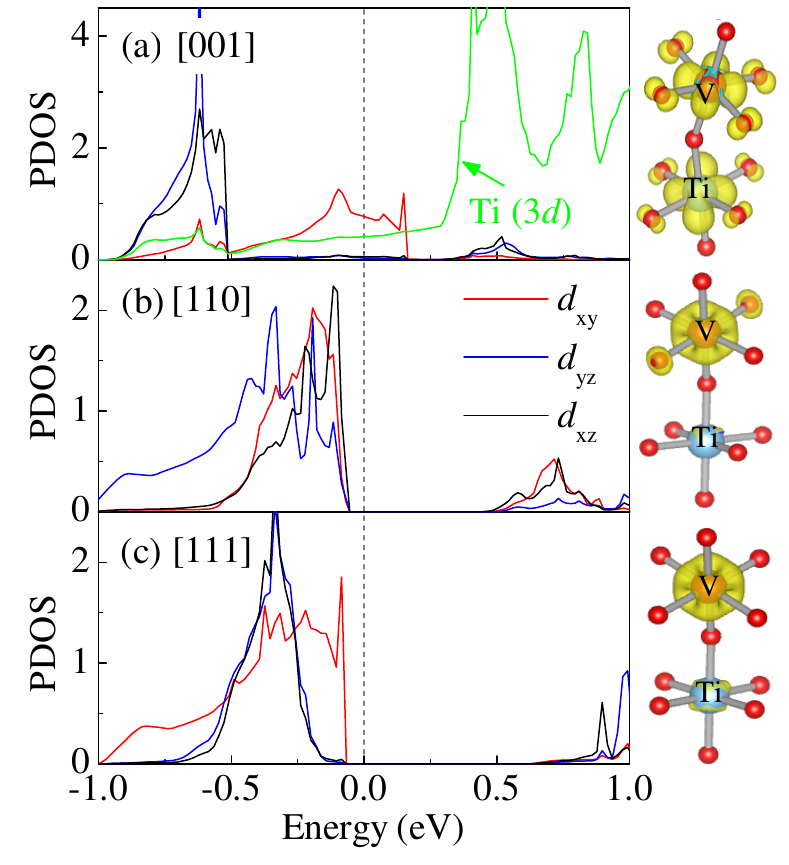}
\vskip -0.7cm
\includegraphics[width=0.45\textwidth]{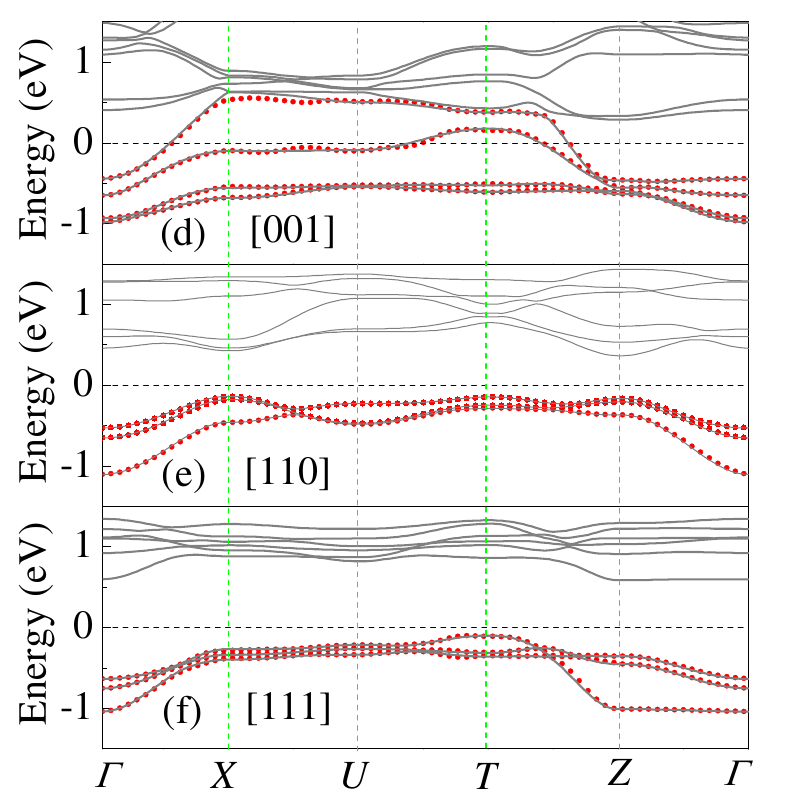}
\vskip -0.3cm
\includegraphics[width=0.43\textwidth]{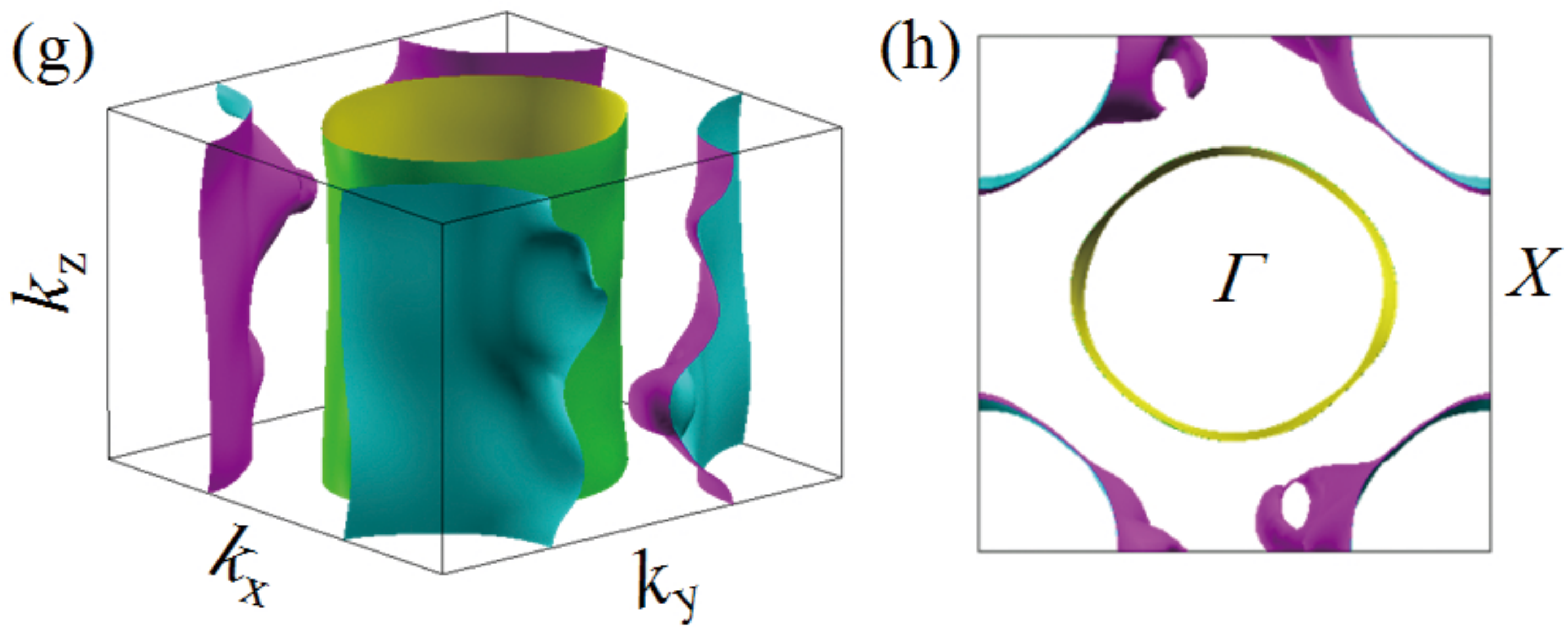}
\caption{The orbital-resolved PDOS around the Fermi level and corresponding spatial distribution of the electron density for the cases of (a) [001], (b) [110], and (c) [111]. (d), (e) and (f) are the Wannier-interpolated band structures (red solid circles) $vs$ original DFT band structure (gray solid lines). Here, the high-symmetric path $\varGamma$-$X$-$U$-$T$-$Z$-$\varGamma$ corresponds to ($0$, $0$, $0$)-($0$, $0.5$, $0$)-($0$, $0.5$, $0.5$)-($0.5$, $0.5$, $0.5$)-($0$, $0$, $0.5$)-($0$, $0$, $0$). The calculated Fermi surface for the [001] case: (g) side view; (h) top view.}
\label{F3}
\end{figure}

In perovskites, the $d$ orbital preference often strongly determines the electronic structures and thus the physical properties, as analyzed in Fig.~\ref{F3}. For the [110] and [111] ones, due to the complete charge transfer, all three $t_{\rm 2g}$ orbitals of V are fully occupied and the corresponding electron clouds own the cubic symmetry. In contrast, for the [001] one, the $d_{xy}$ orbitals of both V and Ti are partially occupied, which can be visualized by the spatial distribution of electron. These partially-filled $d_{xy}$ orbitals allow electrons to move in the (001) Ti-O-Ti and V-O-V planes, leading to two wide bands of $d_{xy}$ orbitals. Such quasi-two-dimensional metallicity can be visualized by plotting the (MLWFs) Fermi surfaces, as shown in Fig.~\ref{F3}(g-h), which are cylindrical with an electronic pocket (from Ti) surrounding the $\varGamma$ point and a hole pocket (from V) around the Brillouin corner. In other words, two-dimensional-conductive electron gas is form, although it is not spatially limited in two-dimensional sheets.

\begin{table}
\caption{MLWFs fitted potential energies (in unit of meV) for V's $t_{\rm 2g}$ orbitals. In each case, the potential of $d_{xy}$-like orbital is set as the energy reference.}
\begin{tabular*}{0.48\textwidth}{@{\extracolsep{\fill}}cccc}
\hline \hline
 & [001] & [110] & [111]\\
\hline
$d_{xz}$-like & $-494$ & $222$    & $23$\\
$d_{yz}$-like & $-535$ & $226$ & $42$\\
$d_{xy}$-like & $0$   & $0$    & $0$\\
\hline \hline
\end{tabular*}
\label{table2}
\end{table}

Using the MLWFs, the on-site potential energies for V's $t_{\rm 2g}$ (spin-up) orbitals are extracted from the DFT bands, as summarized in Table~\ref{table2}. Such on-site potential energies can reflect the overall symmetry and degeneration of triplet $t_{\rm 2g}$ orbitals. For the [001]-oriented case, the Jahn-Teller splitting energy reaches up to $\sim0.5$ eV between $d_{xy}$ and $d_{xz}$/$d_{yz}$. Opposite effect is observed in the [110] case, while the [111] case is more close to high symmetric limit.

\subsection{$2+2$ superlattice}

\begin{figure}
\centering
\includegraphics[width=0.4\textwidth]{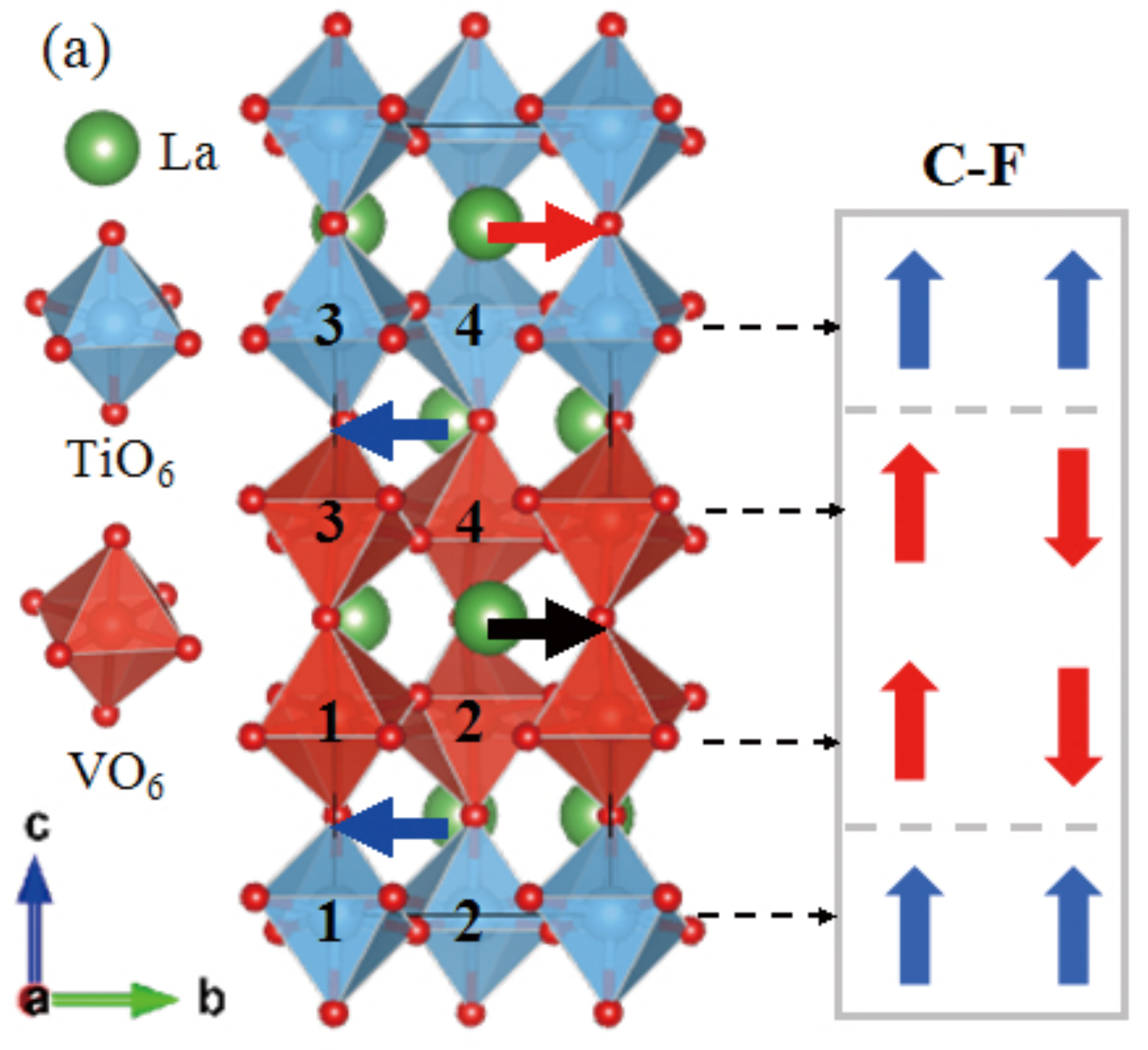}
\includegraphics[width=0.46\textwidth]{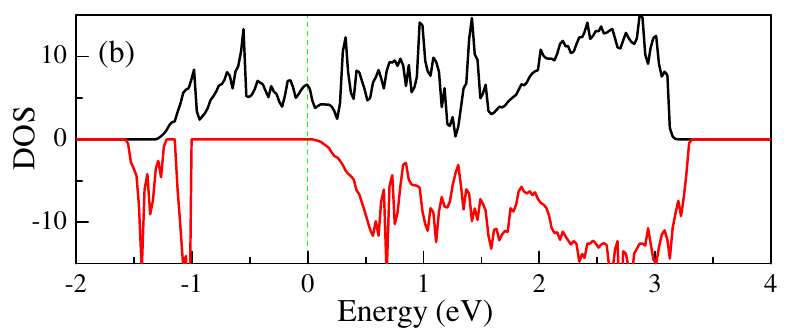}
\includegraphics[width=0.48\textwidth]{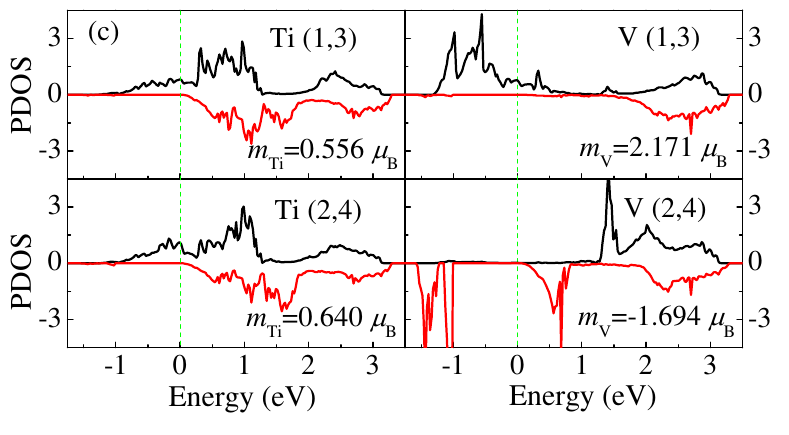}
\caption{(a) Left: Crystalline structure of (LaTiO$_3$)$_2$/(LaVO$_3$)$_2$ superlattice. The uncompensated displacements of A-site ions (La$^{3+}$) is sketched as arrows, which lead to the polar structure. Here, the B-site atoms are labelled from $1$ to $4$. Right: Sketch of the C-F magnetic order. (b) Total DOS. (c) The PDOS for Ti and V sites. The local moments are indicated.}
\label{F4}
\end{figure}

Besides the structural orientation, the thickness periods can also significantly tune the physical properties of superlattices. With increasing thickness, the reduced interface/volume ratio will suppress the interfacial reconstructions, e.g., charge transfer, etc. For example, previous studies revealed the metal-insulator transition in (LaMnO$_3$)$_{2n}$/(SrMnO$_3$)$_n$ superlattices with increasing $n$ \cite{Bhattacharya:Prl,Aruta:Prb}, as well as tuning of exchange bias in (LaMnO$_3$)$_n$/(LaNiO$_3$)$_n$ superlattices \cite{Gibert:Nm}, both of which were related to the change of charge transfer \cite{Dong:Prb08.3,Dong:Prb13.1,Nanda:Prb09}. Thus, it is necessary to go beyond the $1+1$ superlattice, and here the $2+2$ LaTiO$_3$/LaVO$_3$ superlattice grown along the [001] direction will be studied.

\begin{table}
\caption{Summary of DFT energies (in unit of meV) for the $n=2$ superlattice ($40$ atoms). The C-F magnetic state is taken as the reference state for energy comparison. Here, the A-A, C-A, F-F and C-F are hybrid magnetic orders. The A-A (or F-F) state is A-AFM (or FM) order in V's layers and Ti's layers, but coupled ferromagnetically (antiferromagnetically) between V's layer and Ti's layer. The F-A (or C-F) state is FM (or C-AFM) order in V's layers but A-AFM (or FM) order in Ti's layers.}
\begin{tabular*}{0.48\textwidth}{@{\extracolsep{\fill}}ccccc}
\hline \hline
Magnetic order & A-AFM & C-AFM & G-AFM & FM\\
\hline
$\Delta E$ & $98$ & $266$ & $410$ & $163$\\
\hline
Magnetic order & A-A & F-A & C-F & F-F\\
\hline
$\Delta E$ & $153$ & $183$ & $0$ & $96$\\
\hline \hline
\end{tabular*}
\label{table3}
\end{table}


To determine the electronic structure of (LaTiO$_3$)$_2$/(LaVO$_3$)$_2$ superlattice, several possible magnetic orders have been tested, as shown in Table~\ref{table3}. A special hybrid magnetic order with C-AFM order in V's layers but FM order in Ti's layers (denoted as C-F here, see Fig.~\ref{F4}(a)) owns the lowest energy, due to the charge transfer. All tested magnetic orders give robust metallic behavior, not limited to the particular magnetic state.

The total DOS and PDOS of the C-F state are shown in Fig.~\ref{F4}(b-c). The system is a half metal, namely only the spin-up electrons from both Ti and V present around the Fermi level. Similar to the [001]-oriented (LaTiO$_3$)$_1$/(LaVO$_3$)$_1$ superlattice, such a metallic behavior is due to the partial charge transfer. Due to the reduced interface/volume ratio comparing with the $n=1$ case, such partial charge transfer is relative weaker, evidenced by relative larger (smaller) local magnetic moments of Ti (V), as indicated in Fig.~\ref{F4}(c).

In perovskite superlattices with particular stacking modes, e.g., some [001]-stacking ($ABX_3$)$_n$/($A'BX_3$)$_n$ with odd $n$, may break the inversion symmetry and thus induce a polar structure, which have been theoretical formulated and experimentally confirmed \cite{Alaria:Cs14,Bousquet:Nat,Rondinelli:Am12}. Our previous study extended this type of polar structure to the [001]-stacking ($ABX_3$)$_n$/($AB'X_3$)$_n$ with even $n$ \cite{Zhang:Prb15}.

Indeed, here the space group of [001]-stacking (LaTiO$_3$)$_2$/(LaVO$_3$)$_2$ superlattice is $Pmc2_1$ and its corresponding point group is $mm2$, which is a polar point group. The origin of polar structure can be visualized in Fig.~\ref{F4}(a). Similar to (YTiO$_3$)$_2$/(YFeO$_3$)$_2$ superlattice \cite{Zhang:Prb15}, both La$^{3+}$ cations and O$^{2-}$ anions are move away from their corresponding high-symmetric positions due to the octahedral tilting. And the sequence of B-site cations along the $c$ axis, i.e., ...-Ti-Ti-V-V-..., would break the space inversion symmetry \cite{Dong:Prl14,Dong:Ap}. Thus, the [001]-stacking (LaTiO$_3$)$_2$/(LaVO$_3$)$_2$ should also be robustly polar, protected by the particular geometric structure.

There is no doubt that the multifunctional materials with unusual coexisting properties, e.g. polar structure and metallicity, could broaden the new research area of electronic devices. In fact, the polar metals have been observed in LiOsO$_3$ \cite{Shi:Nm,Puggioni:Prl} and thin-film $R$NiO$_3$ \cite{kim:Nat}, and also predicted in ruthenate oxide \cite{Puggioni:Nc}. In our previous studied (YFeO$_3$)$_2$/(YTiO$_3$)$_2$, the metallicity is not robust, or even artificial. Here both the noncentrosymmetric polar structure and metallicity are more reliable, which need further experimental confirmation/verification.

\subsection{Additional discussion}
In oxide heterostructures, both the strain effect and charge transfer can tune the physical properties, as demonstrated above. Generally, the strain effect modifies the octahedral shape and tunes the energy levels of $3d$ orbitals. For correlated electrons, the tuning of orbital levels may lead to the change of magnetic ground state as well as metallicity, since both the exchange interactions and band structures are seriously depend on the orbital occupancy. The charge transfer can also seriously affect the magnetism and electronic structure. Thus, these two effects may be entangled and cooperating together. Sometimes it is not easy to distinguish the individual effects of these two variables.

In the current study, for the [110]- and [111]-oriented $n=1$ superlattices, due to the full charge transfer from Ti to V, the epitaxial strain effects on these two materials can be neglected, considering Ti's $d^0$ and V's $d^3$ configurations. In contrast, for the [001]-oriented $n=1$ and $n=2$ ones, due to the partial charge transfer, the orbital degree of freedom is still active. In this case, the electronic structures are sensitive to the strain.


\begin{figure}
\centering
\includegraphics[width=0.47\textwidth]{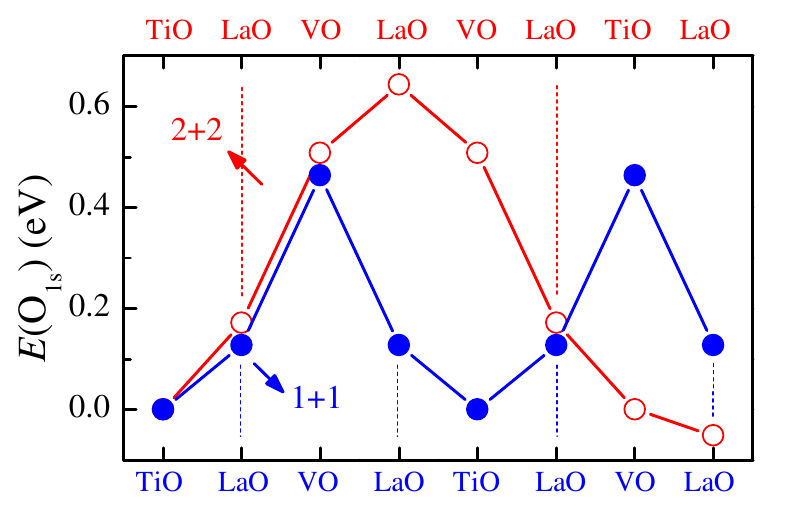}
\caption{Modulation of oxygen's $1s$ core-level energy in each LaO/TiO$_2$/VO$_2$ layers of the [001]-oriented superlattices. Solid (blue): $n=1$ and open (red): $n=2$. The interfaces, i.e., the LaO layers sandwiched between TiO$_2$ and VO$_2$ layers, are indicated by dashed lines.}
\label{F5}
\end{figure}

Due to the charge transfer, the electrostatic potential will be modulated through the superlattices, which will react to the charge transfer and finally reach a balance between charge transfer and electrostatic potential. In DFT, this potential modulation can be characterized using some deep-energy levels, e.g. O's $1s$ core level ($E({\rm O}_{1s})$) \cite{Nanda:Prb09,Zhang:Prb12}, as shown in Fig.~\ref{F5} for the [001]-stacking superlattices. The Ti (V) sites own lower (higher) electrostatic potential, as expected from the charge transfer direction. The offset between Ti site and V site is only slightly larger in the $n=2$ case, since all TiO$_2$ and VO$_2$ layers are interfacial layers (thus no inner layer) in both the $n=1$ and $n=2$ cases. Even though, it is clear that the inner LaO layer in LaVO$_3$ region own much higher potential, thus it is natural to expect that the potential barriers between Ti and V will be apparently increased in larger $n$ cases, similar to the (LaMnO$_3$)$_{2n}$/(SrMnO$_3$)$_n$ superlattices \cite{Nanda:Prb09,Zhang:Prb12}.

\section{Conclusion}
In summary, the (LaTiO$_3$)$_n$/(LaVO$_3$)$_n$ superlattices have been studied using the first-principles calculation. As a model system, many interesting topics of interfacial physics emerge in these heterostructures, including the unconventional charge transfer contrary to the intuitional band alignment, orbital ordering related metal-insulator transitions, as well as the polar metallicity. Further experimental studies are expected to confirm/verify these predictions.

\begin{acknowledgments}
Work was supported by the National Natural Science Foundation of China (Grant No. 11674055), Jiangsu Innovation Projects for Graduate Student (Grant No. KYLX15\underline{ }0112), and the Scientific Research Foundation of Graduate School of Southeast University (Grant No. YBJJ1619). Most calculations were done on Tianhe-2 at National Supercomputer Center in Guangzhou (NSCC-GZ).
\end{acknowledgments}

\bibliographystyle{apsrev4-1}
\bibliography{ref}
\end{document}